\documentclass[journal=jacsat,manuscript=article,layout=twocolumn]{achemso}
\usepackage{chemformula} 
\usepackage{color,soul}
\usepackage[T1]{fontenc} 
\usepackage{amsmath,amssymb}

\usepackage{tabularx}
\usepackage{array}
\usepackage{dcolumn}
\usepackage{setspace}

\author{Vasily S. Stolyarov}
\email{vasiliy.stoliarov@gmail.com}
\affiliation{Advanced mesoscience and nanotechnology centre, Moscow Institute of Physics and Technology, 141700 Dolgoprudny, Russia}\alsoaffiliation{Dukhov Research Institute of Automatics (VNIIA), 127055 Moscow, Russia}
\alsoaffiliation{National University of Science and Technology MISIS, 119049 Moscow, Russia}

\author{Vsevolod~Ruzhitskiy}
\affiliation{Dukhov Research Institute of Automatics (VNIIA), 127055 Moscow, Russia}

\author{Razmik~A.~Hovhannisyan}
\affiliation{Advanced mesoscience and nanotechnology centre, Moscow Institute of Physics and Technology, 141700 Dolgoprudny, Russia}

\author{Sergey~Yu.~Grebenchuk}
\affiliation{Advanced mesoscience and nanotechnology centre, Moscow Institute of Physics and Technology, 141700 Dolgoprudny, Russia}

\author{Andrey~G.~Shishkin}
\affiliation{Advanced mesoscience and nanotechnology centre, Moscow Institute of Physics and Technology, 141700 Dolgoprudny, Russia}
\alsoaffiliation{Dukhov Research Institute of Automatics (VNIIA), 127055 Moscow, Russia}

\author{Igor~A.~Golovchanskiy}
\affiliation{Advanced mesoscience and nanotechnology centre, Moscow Institute of Physics and Technology, 141700 Dolgoprudny, Russia}
\alsoaffiliation{National University of Science and Technology MISIS, 119049 Moscow, Russia}
\alsoaffiliation{Dukhov Research Institute of Automatics (VNIIA), 127055 Moscow, Russia}

\author{Alexander A. Golubov}
\affiliation{Faculty of Science and Technology and MESA+ Institute of Nanotechnology, 7500 AE Enschede, The Netherlands}

\author{Nikolay~V.~Klenov}
\affiliation{Faculty of Physics, Lomonosov Moscow State University, Moscow, 119991, Russia}
\alsoaffiliation{Dukhov Research Institute of Automatics (VNIIA), 127055 Moscow, Russia}

\author{Igor~I.~Soloviev}
\affiliation{Skobeltsyn Institute of Nuclear Physics, Lomonosov Moscow State University, Moscow, 119991, Russia}
\alsoaffiliation{Dukhov Research Institute of Automatics (VNIIA), 127055 Moscow, Russia}

\author{Mikhail~Yu.~Kupriyanov}
\affiliation{Skobeltsyn Institute of Nuclear Physics, Lomonosov Moscow State University, Moscow, 119991, Russia}

\author{Alexander V. Andriyash}
\affiliation{Dukhov Research Institute of Automatics (VNIIA), 127055 Moscow, Russia}

\author{Dimitri Roditchev}
\affiliation{LPEM, ESPCI Paris, PSL Research University, CNRS, 75005 Paris, France}
\alsoaffiliation{Sorbonne Universite, CNRS, LPEM, 75005 Paris, France}

\title[An \textsf{achemso} demo]{Revealing Josephson vortex dynamics in proximity junctions below critical current}

\begin{document}

\begin{figure}[ht!]
\begin{center}
\includegraphics[width=7 cm]{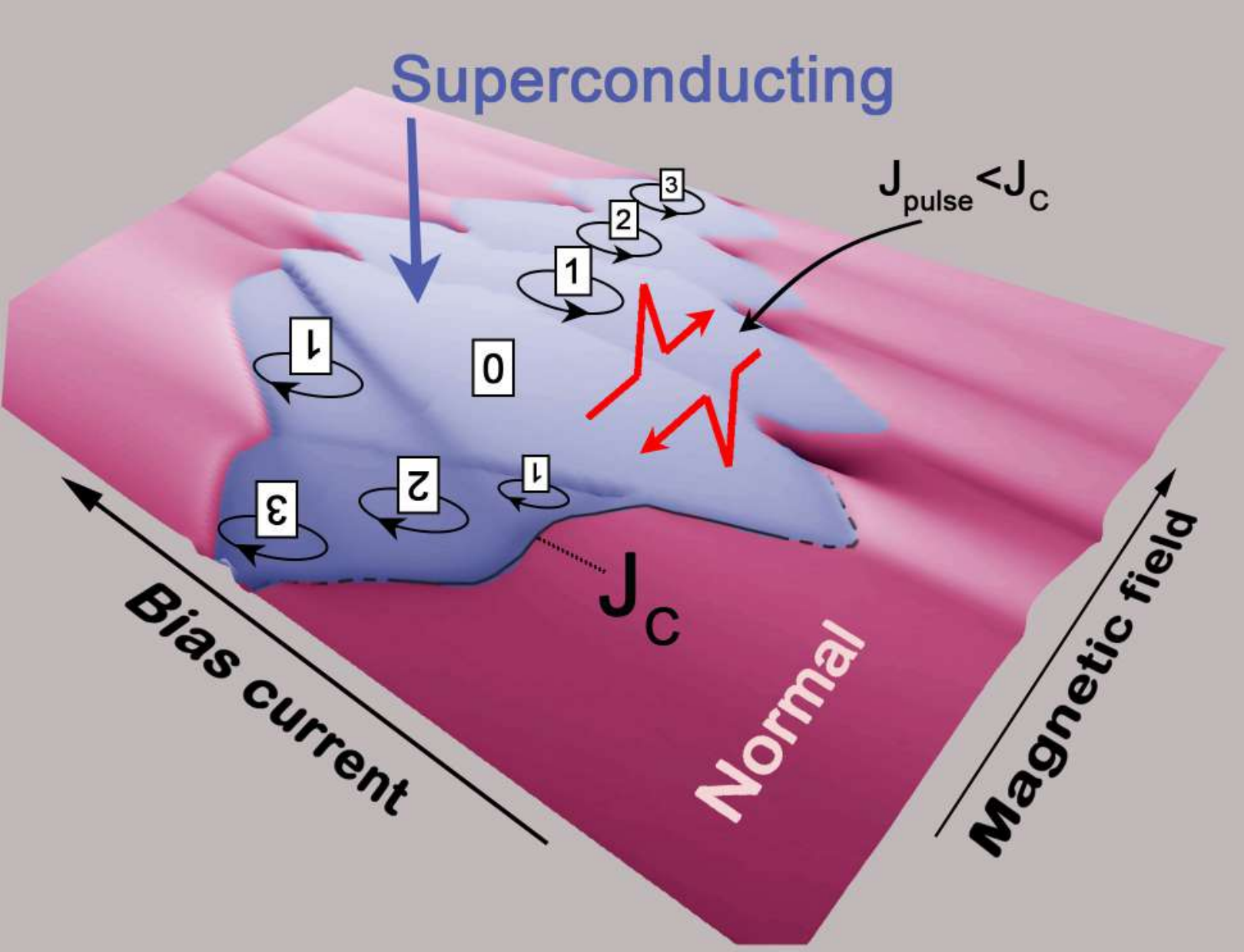}
\end{center}
\end{figure}

\begin{abstract}
Made of a thin non-superconducting metal (N) sandwiched by two superconductors (S),  SNS Josephson junctions enable novel quantum functionalities by mixing up the intrinsic electronic properties of N with the superconducting correlations induced from S by proximity. Electronic properties of these devices are governed by Andreev quasiparticles \cite{Andreev} which are absent in  conventional SIS junctions whose insulating barrier (I) between the two S electrodes owns no electronic states. Here we focus on the Josephson vortex (JV) motion inside Nb-Cu-Nb proximity junctions subject to electric currents and magnetic fields. The results of local (Magnetic Force Microscopy) and global (transport) experiments provided simultaneously are compared with our numerical model, revealing the existence of several distinct dynamic regimes of the JV motion. One of them, identified as a fast hysteretic entry/escape below the critical value of Josephson current, is analyzed and suggested for low-dissipative logic and memory elements.
\end{abstract}


\section{Introduction}

\begin{figure*}[ht!]
\begin{center}
\includegraphics[width=17cm]{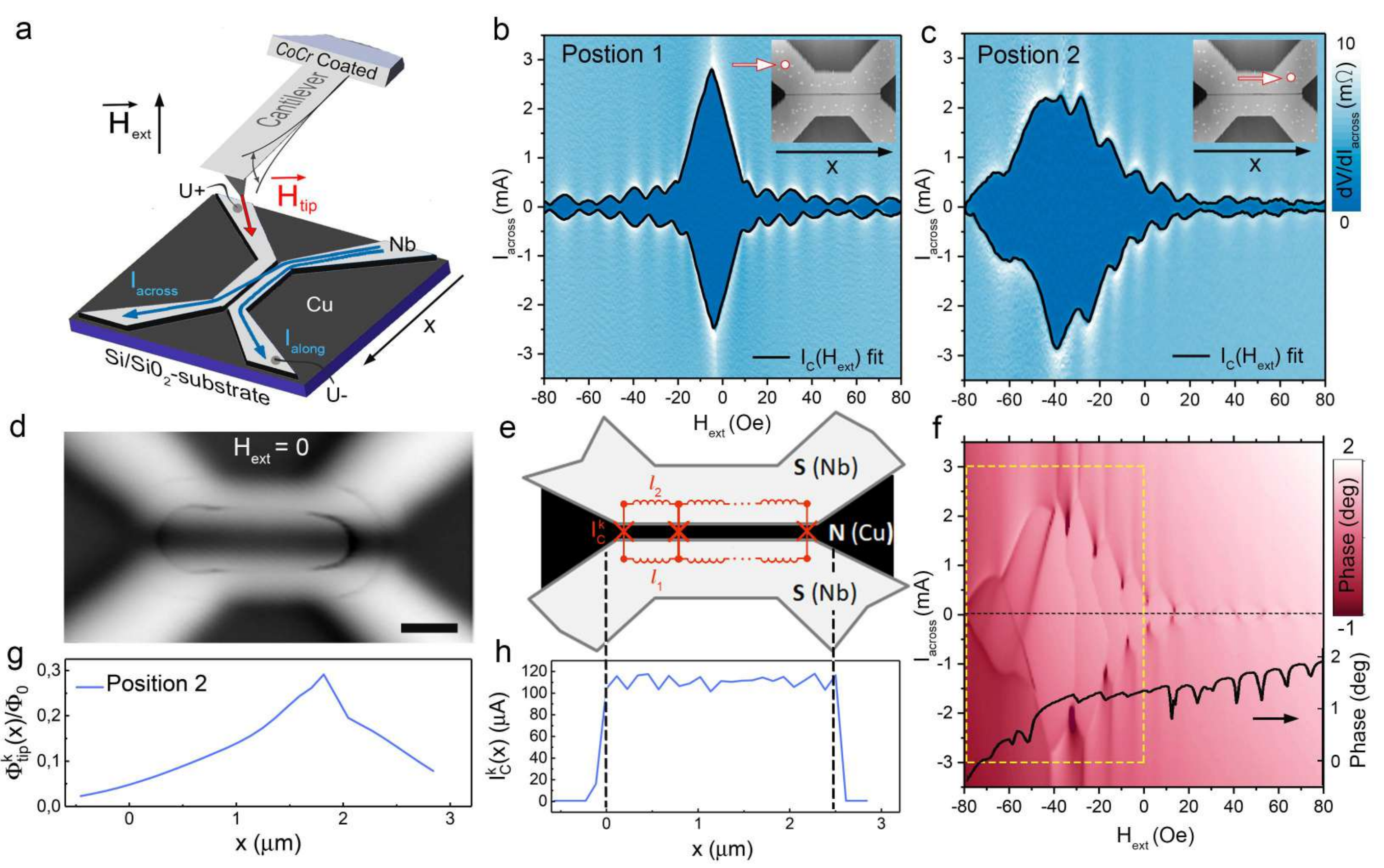}
\caption{\textbf{a} - Experimental setup: 100 nm thick Nb leads (in light gray) are patterned on a 50 nm thick Cu layer (in dark gray) on Si/SiO$_2$ substrate (in dark blue).The junction area between Nb-leads is 2500 nm x 200 nm. The transport properties are measured by applying current I$_{across}$ (called $I$ below) across Nb-leads and by measuring the voltage drop $V=$ U+ - U-. The external magnetic flux inside the junction is due to an external field $H_{ext}$ and the field $H_{tip}$ of the oscillating Co/Cr-coated tip of a MFM; an additional flux is created by applying supercurrent I$_{along}$ along superconducting leads. \textbf{b,c}~-~color-coded $dV/dI(I,H_{ext})$ maps of the device at different tips locations (arrows in insets show positions 1 and 2, the respective tip heights are 2000 and 70~nm). Dark-blue: zero-resistance $V=0$ regime of the junction; light-blue: resistive regime $|V|>0$;  black solid lines are fits by RSJ model (see in the text); \textbf{d} - spatial tip oscillation phase map of the sample (T = 4.5~K, $H_{ext}$=0). Black arcs are  locations where tip triggers JV entries. The scale bar corresponds to 0.5~$\mu$m. \textbf{e} - schematic top view of the device with the equivalent circuit (in red) used for discrete RSJ modeling (see in the text). \textbf{f} - color-coded tip oscillation phase map $Phase(I,H_{ext})$ at the tip position 2. Black curve - a $Phase(H_{ext})$ profile at $I$=0. \textbf{g,h} - the tip-induced flux and the critical current distributions required to obtain the RSJ fit in panel (c).}
\label{fig:scheme}
\end{center}
\end{figure*}

The phase portraits of quantum coherent condensates - superconductors, superfluids, cold atoms and ions - may contain 2$\pi$-phase loops known as quantum vortices \cite{Abrikosov_1957,Stolyarov_2018}. Such loops may also appear in Josephson junctions (JJ). These Josephson vortices (JV) \cite{Josephson_1962,Anderson_1963,Rowell_1963,Josephson_1965,Wallraff_2003,Berdiyorov_2018} were described\cite{Cuevas_2007} and recently observed inside normal parts of SNS JJs \cite{Cuevas_2007,Roditchev_2015,Dremov_2019,Grbenchuk_2020,Hovhannisyan_2021}. The integer number $n$ of Josephson vortices present in a JJ is associated with the $n$-th branch of the critical current modulation vs magnetic field~ $I_c(H)$\cite{Grbenchuk_2020,Krasnov_2020}.

Generation and manipulation of JVs \cite{Malomed_2004, Gulevich_2006, Gulevich_2017, Knufinke_2012} is a basis of many applications of superconducting technology. This includes quantum computing based on flux \cite{FlQ1,FlQ2} and JV \cite{JVQ1,JVQ2,Wallraff_2003} qubits, control of quantum circuits \cite{QuProcMukh,SolScat,BalRoc}, novel superconducting neural networks with information encoded in the magnetic flux \cite{SCNN1,SCNN2,SCNN3}, reservoir computing based on superconducting electronics \cite{ResC}, superconducting digital and mixed-signal circuits \cite{Holmes,Mukh2,Tolp,Beil,IRDS2020}, cryogenic memory~\cite{Semenov_2019, Ilin_2021, Miloshevsky_2022, Ligato_2021}.

One of the advantages of superconducting devices is related to their  low power dissipation. The low-dissipative dynamics of JVs is a highly desired mode of operation where, in the limiting case, the JV is a subject of adiabatic process characterized by nearly zero energy exchange with the environment.
However, the detection and manipulation of individual JV, the study of their low-dissipative motion is an experimental challenge, especially in long Josephson junctions where the nucleation of JV in one part of the junction does not necessarily lead to the appearance of a measurable voltage drop across the entire junction. This is due to the compensating redistribution of the screening currents throughout the superconducting device.

\begin{figure*}[t!]
\begin{center}
\includegraphics[width=12cm]{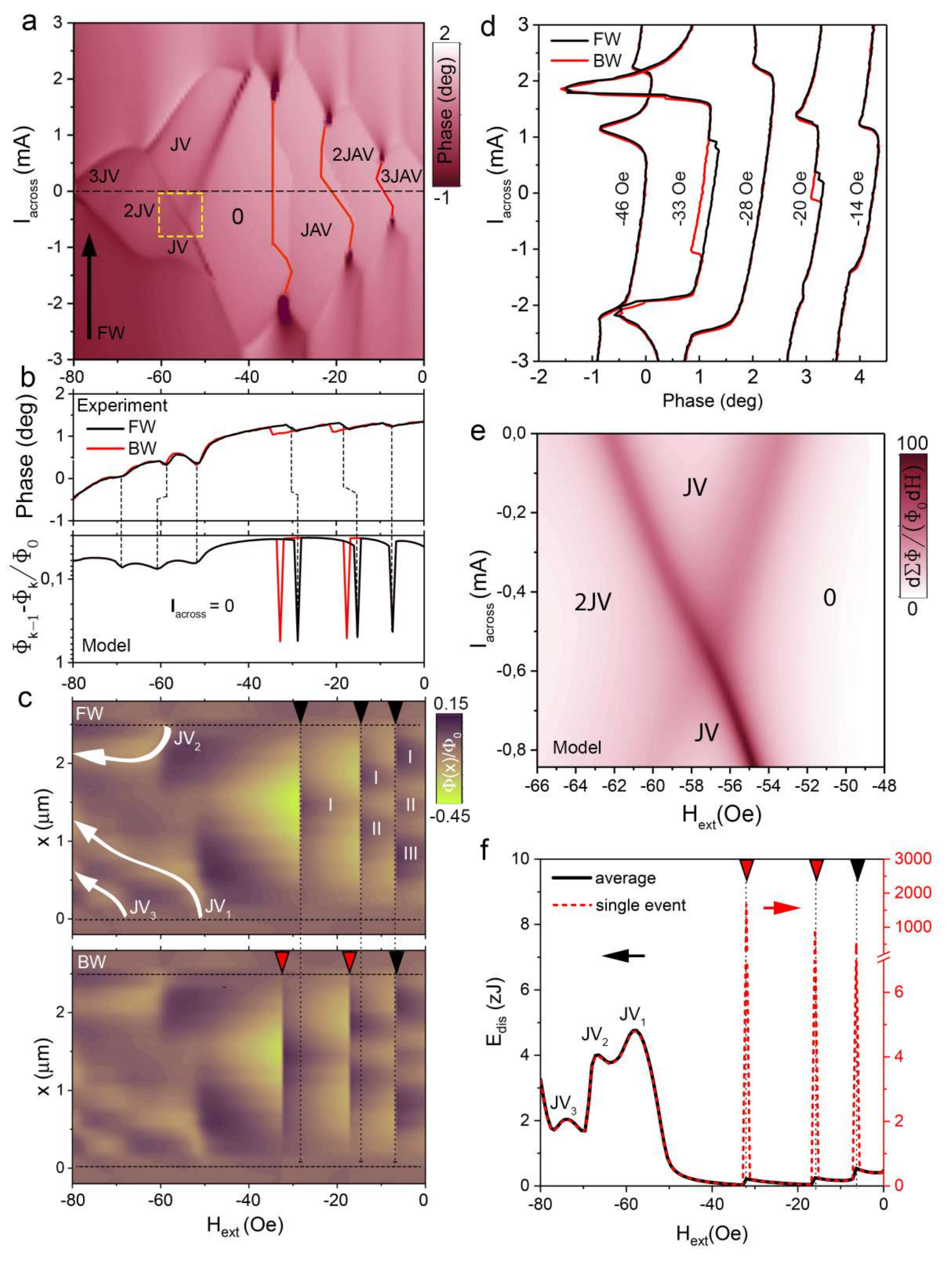}
\caption{\textbf{a}~-~a zoomed view on the phase map $Phase(I,H_{ext})$ of Fig. 1(f) (forward field sweep). Labels denote Josephson vortices and Josephson anti-vortices present inside the device below $I_c$. At the backward field sweep some phase boundaries are shifted (red lines). Dashed square: region of interest and of the fit in Fig. 2(e) (see in the text). \textbf{b}~-~upper panel: $Phase(I=0,H_{ext})$ plots for forward (black curves) and backward (red curves) sweeps. Lower panel: field dependence of the corresponding total magnetic flux (calculation, see in the text). \textbf{c}~-~color-coded magnetic flux spatial distribution inside the junction during forward and backward field sweeps (calculation, see in the text). Horizontal lines: junction edges. White labels and arrows show the evolution of individual JV positions. \textbf{d}~-~$Phase(I)$ cuts at selected fields $H_{ext}$ for forward and backward sweeps. \textbf{e}~-~simulation of the crossing point in Fig. 2(a) (see in the text). \textbf{f}~-~magnetic field dependence (backward sweep) of the energy dissipated in the junction over one tip oscillation period; red dashed line - for the first oscillation after $H_{ext}$ change, black line -  averaged over large number of following oscillations (see explanations in the text).} 
\label{fig:Fraun}
\end{center}
\end{figure*}

In this paper, we study the dynamics of Josephson vortices in lateral Nb-Cu-Nb Josephson junctions. The JVs evolve in an inhomogeneous magnetic field $H=H_{ext}+H_{tip}(r,t)$, where $H_{ext}$ is a homogeneous field produced by an external superconducting coil, and $H_{tip}$ is a spatially inhomogeneous field  generated by oscillating magnetic tip of the magnetic force microscope (MFM), Fig.~\ref{fig:scheme}(a). Experimentally, the JVs are detected simultaneously in the current-voltage $V(I)$ characteristics of the device, Figs.~\ref{fig:scheme}(b,c), and in the phase of the MFM tip resonant oscillations, Fig.~\ref{fig:scheme}(d,f). The numerical RSJ model we present here was first validated to perfectly fit the experimentally established current-field characteristics, $I_c(H_{ext})$ \cite{Ruzhitskiy_2021}
. It was further used to restore fast JV dynamics, even in cases $I<I_c$ when no voltage drop across the junction could be experimentally detected. Finally, we quantify the speed and the energy dissipation during the JV motion, and discover the existence of different dynamic regimes characterized by JV dissipation varying over 3 orders of magnitude.

\section{Experiment and Modelling}

The sketch of the studied device is presented in Fig.~\ref{fig:scheme}(a). The JJ consist of two 100 nm thick Nb-leads patterned on 50 nm thick Cu-film. The gap between Nb-electrodes is 200~nm, the junction width is 2500~nm. Outside the junction area, Nb-leads are made larger, to do not limit the critical current of the junction.

The voltage-current-field $V(I,H)$ transport characteristics were recorded by applying the electric current across the junction (noted I$_{across}$ in Fig.~\ref{fig:scheme}(a)),  and measuring the voltage drop between the electrodes U+ and U-. The magnetic field was applied perpendicularly to the sample surface. The critical temperature of the superconducting transition of Nb-electrodes was 7.2~K, the critical Josephson current $I_c = 2.8$~mA $@$ 4.2 K. The MFM experiments were provided at 4.5 K using a commercial Attocube attoDRY 1000 microscope with Co-coated magnetic cantilevers. Further experimental details are available elsewhere \cite{Dremov_2019,Grbenchuk_2020,Hovhannisyan_2021}.

As expected, the penetration of individual JVs into the junction is reflected in Fraunhofer-like oscillations of the critical Josephson current $I_c(H)$. In Fig.~\ref{fig:scheme}(b,c), the voltage across the junction is presented in false color as a function of $I$ and $H_{ext}$. On these maps, the superconducting regions $|I|<I_c$ correspond to zero-voltage drops and appear in dark blue. The conditions for the $n$-th JV entry depend on both the external field  $H_{ext}$ and the field $H_{tip}(r,t)$ of oscillating MFM tip. Consequently, Fraunhofer-like oscillations get distorted, depending on tip position over the JJ, and $V(I,H_{ext})$ maps vary significantly (compare Fig.~\ref{fig:scheme}(b) and (c) along with the  tip positions presented in respective insets). Indeed, the magnetized tip produces at the JJ a field  of several tens of Oe \cite{Hovhannisyan_2021}, that is strong enough to provoke the entry of several JVs, even at $H_{ext}=0$. The entries of individual JVs are detected as sudden phase drops of the tip oscillations, appearing in MFM maps, Fig.~\ref{fig:scheme}(d), as black arcs  \cite{Dremov_2019}.

The phase signal contains therefore a precious information about JV penetration events which are sometimes inaccessible in the transport data. This ability is offered by the magnetic interaction between the JJ and the out-of-plane oscillating MFM tip. The latter produces at JJ a field $H_{tip}(r,t)=H_{tip}(r)+\delta H_{tip}(r,t)$, where $H_{tip}(r)$ is a stationary  and $\delta H_{tip}(r,t)<<H_{tip}(r)$ a tiny oscillatory parts. When the JJ is in a stable JV configuration, the field $\delta H_{tip}(r,t)$ provokes only a slight cyclic motion of JVs inside JJ around their equilibrium positions. But when JJ is driven very close to a transition between two different JV configurations (by applying $H_{ext}$ or a current $I_{along}$ along one of the electrodes), the oscillatory field $\delta H_{tip}(x,t)$ can be enough to trigger the JV entry/escape
. In turn, the dissipation due to the JV motion affects the tip oscillation and appears as sudden phase drops  \cite{Dremov_2019,Grbenchuk_2020,Hovhannisyan_2021}. That is why $V(I,H_{ext})$ map in Fig.~\ref{fig:scheme}(c) and the simultaneously recorded phase map $Phase(I,H_{ext})$ in Fig.~\ref{fig:scheme}(f) resemble each other. Though, beyond similarities, the phase map reveals some events occurring inside $V=0$ region, that is when the junction remains globally in the superconducting state, $|I|<I_c$.  A zoom on the corresponding phase map region of interest (dashed yellow rectangle) is presented in Fig.~\ref{fig:Fraun}(a) and discussed later.

The DC-transport and MFM data provide only a quasi-static picture of the junction behaviour. To account for the JV dynamics, we simulate the junction behaviour using the discrete sine-Gordon equation approach \cite{Golovchanskiy_SUST_30_054005}. The junction is divided in $k=$31 elementary junctions, see Fig.~\ref{fig:scheme}(e). The low capacitance of the junction is neglected in calculations. 

\begin{figure*}[ht!]
\begin{center}
\includegraphics[width=18cm]{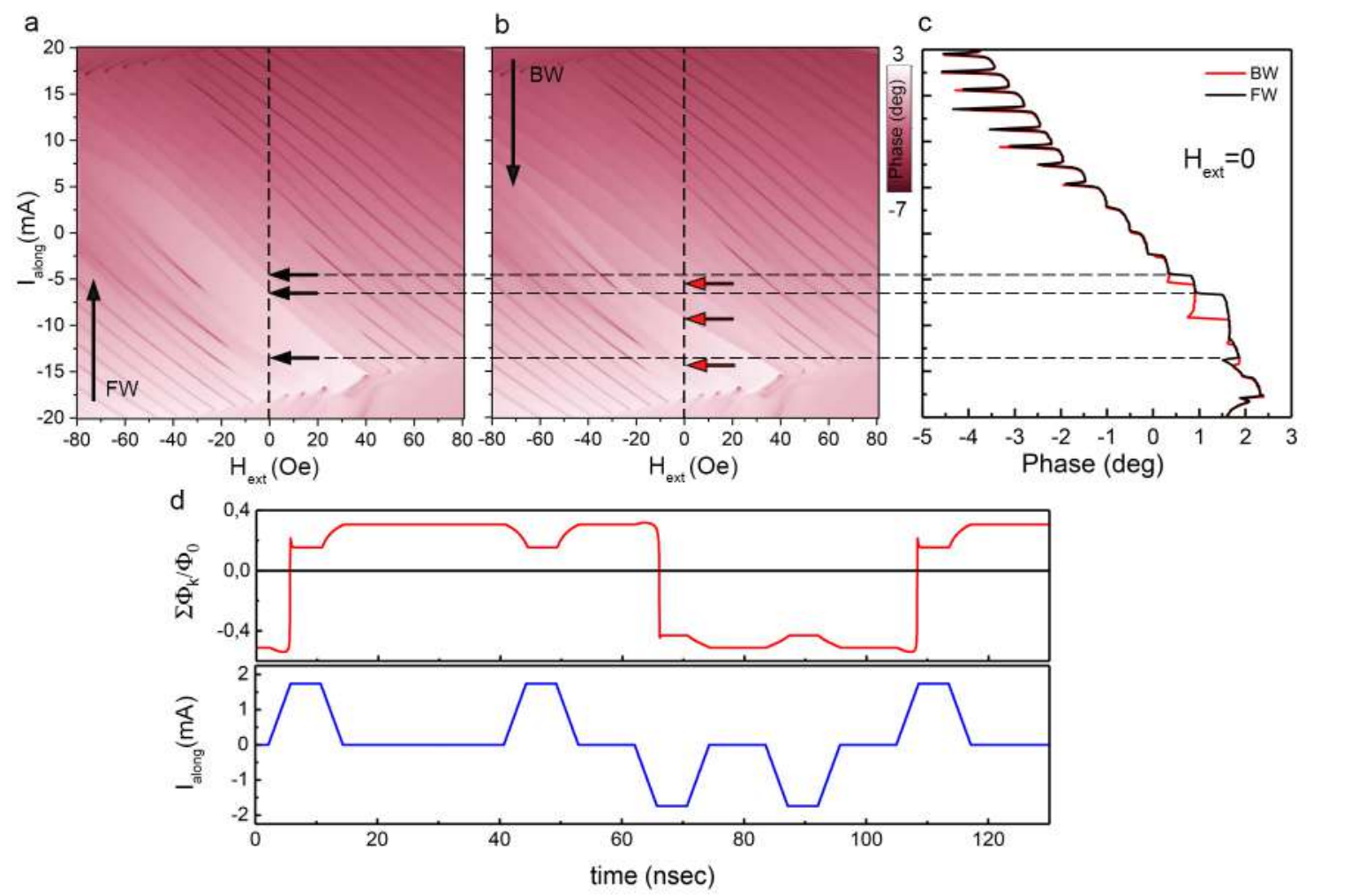}
\caption{\textbf{a,b}~-~color-coded tip oscillation phase $Phase$ as a function of magnetic field $H_{ext}$ and current $I_{along}$ for forward, in (a), and backward, in (b), $I_{along}$ sweeps. \textbf{c}~-~$Phase(I_{along},H_{ext}=0)$ cuts of the maps (a) and (b). \textbf{d}~-~Driving current pulse sequence $I_{along}(t)$ (lower panel) and corresponding evolution of the total magnetic flux inside the junction at $H_{ext}=-$33 Oe (upper panel) demonstrating the memory behaviour (see in the text). }
\label{IPH-along}
\end{center}
\end{figure*}

At the first step of simulation, we calculated $I_c(H_{ext})$ curves for various tip positions. The results for Position 1 and Position 2 are presented as black solid lines superimposed on $V(I,H_{ext})$ maps in Figs.~\ref{fig:scheme}(b),(c); they perfectly match the experimental data. 
The correctness and robustness of the model is not warranted  by the high fit quality only but also by the fact that these fits are obtained using physically reasonable parameters. For instance, the junction surface used for the flux calculation has been taken $A=L\times d$, with $L = 2500$~nm, that is the geometrical length of the LJJ, and $d=$330 nm. The latter value is indeed close to the so-called magnetic thickness of the junction, $d_{eff} = t + 2\lambda \coth(d/\lambda) \simeq 380$ nm, where $t$=200 nm is the geometrical distance between the two Nb-electrodes, $\lambda$=90 nm is the London penetration depth in Nb, and $d$=100 nm is the Nb-film thickness \cite{Hovhannisyan_2021}. The Josephson penetration depth (which is a parameter in the model accounting for the flux screening by N-part of the junction) is found to be $\lambda_J \approx 633$~nm - a very reasonable value for 50 nm thick Cu. The junction length is therefore $3.95 \lambda_J$ that justifies the long JJ regime. The distributions $H_{tip}(x)$ and $I_c(x)$ which are main parameters required to fit $I_c(H_{ext})$ data are presented in Figs.~\ref{fig:scheme}(g),(h), respectively. The $H_{tip}(x)$ has a maximum on the right side of the junction, as expected for the geometrical position of the tip. Its magnitude $\sim$3-30 Oe matches the experimentally established value \cite{Hovhannisyan_2021}. The $I_{c}(x)$ dependence is found flat meaning that all 30 cells contribute almost equally to the total Josephson current. This is indeed expected for a high quality JJ. A rapid decay of $I_{c}(x)$ outside the junction limits is in a good agreement with the junction geometry.  
Finally, a tiny mutual inductance between the feed lines and the junction was considered and estimated to $0.3$~fH. This inductance accounts for a tiny skew of the $I_c(H)$ dependencies such that their main extrema for the opposite current polarities are located at slightly different values of the external magnetic field in Fig.~\ref{fig:scheme}(b).




At the second step of modelling, we go beyond the experiment and restore the magnetic flux distribution in the junction (all parameters found at the first step are kept fixed). We focus on the  superconducting regime $|I|<I_c$ in the field window $H_{ext}$ from 0 to -80 Oe (the negative sign is given with respect to $H_{tip}$ considered positive). This part of the $Phase(I,H_{ext})$ map is most intriguing, where several JV penetration events occur, Fig.~\ref{fig:Fraun}(a). Thin dark lines on the map are the phase drops - the moments of individual JV entries. This enables us identifying all JV phases unambiguously. The region numbered "0" stands for no JV inside the junction, "JV", "2JV", 3JV" - for one, two and three Josephson vortices whose flux direction  is aligned with $H_{ext}$, and "JAV", "2JAV", "3JAV" stand for one, two and three Josephson anti-vortices which are JV of the opposite polarity (aligned with $H_{tip}$). For completeness, the graphs $Phase(H_{ext})$ at $I=0$ are presented in the upper part of Fig.~\ref{fig:Fraun}(b) for both forward (black line) and backward (red line) field sweeps; the cuts $Phase(I)$ at selected values of $H_{ext}$ are displayed in Figs.~\ref{fig:Fraun}(d). These cuts demonstrate that some JV entry/escape events are sharp and hysteretic while others are not.

The $I=0$ regime, presented by horizontal dashed line in Fig.~\ref{fig:Fraun}(a), is of particular interest as, by varying $H_{ext}$, several JV and JAV phases are attended while keeping the device well below $I_c$ (that is at zero DC-voltage across JJ). The corresponding numerically calculated color-coded flux maps are shown in Fig.~\ref{fig:Fraun}(c), the upper map corresponding to the forward field sweep 0 $\rightarrow$~\ -80 Oe, and the lower map - to the backward sweep. In these maps the known flux due to the external field and to the field of the MFM tip is subtracted. The displayed flux is generated by spatially inhomogeneous Josephson currents in N-parts of JJ and by Meissner currents in S-electrodes.  

These flux maps confirm that at $H_{ext}=$0, there are already 3JAV inside the junction. As expected, JAV appear as dark spots in Fig.~\ref{fig:scheme}(g) due to their positive flux; their positions noted I, II and III  are shifted towards the right (top) part of the junction  where the inhomogeneous $H_{tip}(x)$ creates as sort of a potential well for these JAVs.

When a negative $H_{ext}$ is applied, the total flux through the junction is reduced and the 2JAV configuration becomes thermodynamically stable. However, the flux remains spatially inhomogeneous: In the central part of the junction, it is still positive due to $H_{tip}>$0, while at the edges it becomes negative. This creates additional potential barriers for the 3rd JAV to exit through the edges. The barriers render the 3JAV $\rightarrow$ 2JAV transition sharp and hysteretic in field. The same phenomenon is observed for 2JAV $\rightarrow$ JAV and JAV $\rightarrow$ 0 transitions. This numerical result is in a perfect agreement with the experiment (compare experimental and theoretical curves in Fig.~\ref{fig:Fraun}(b)).


At $H_{ext}\approx$~\ -30 Oe, the last JAV escapes the JJ and the system enters an inhomogeneous Meissner state: One can see an extra flux generated by circulating Josephson and Meissner currents trying to screen the total field $H(x)$. At $H_{ext}\approx$~\ -52 Oe, the first JV of the negative polarity enters the junction (labelled JV$_1$ in Fig.~\ref{fig:Fraun}(c)). Repelled from the MFM tip ($H_{tip}>$0), this JV enters from the $x=0$ edge and remains localized there. Since $H_{ext}<$0 does not produce edge barriers for JVs, the transition from Meissner to JV state is smooth and non-hysteretic.  

When $H_{ext}$ is swept further, new JVs enter. The 2nd JV gets in at $H_{ext}\approx$~\ -60 Oe from the right ($x= 2.5 \mu$m) edge: the 3rd JV - at $H_{ext}\approx$~\ -67 Oe from the left ($x= 0$) edge of JJ. These transitions are also smooth and non-hysteretic, as expected in the absence of edge barriers. Note that the penetration of $n$-th JV modifies the location and flux distribution of already present $(n-1)$ JVs.

\begin{singlespace}
\begin{table*}[t!]
\centering
	\newcolumntype{Y}{>{\hsize=0.8\hsize\linewidth=5\hsize\centering\arraybackslash}X}
	\setlength{\extrarowheight}{3pt}
	\begin{tabularx}{\textwidth}{X|X|X|X}
		\multicolumn{4}{l}{TABLE I. Parameters of Josephson vortices: $\tau$ -- vortex penetration time,
		$V$ -- average vortex}\\
        \multicolumn{4}{l}{speed, $E_{dis}$ -- dissipation energy when changing the number of vortices in the junction}\\
		\hline \hline Vortex transition  & Time~$\tau$,~ns & Speed~$V$,~m/s & Energy~$E_{dis}$,~aJ\\
		\hline
		$0JAV \rightarrow 1JAV$ & 0.4 & $4.6  \cdot 10^3$ & 1.7\\
		$1JAV \rightarrow 2JAV$ & 0.7 & $1.7 \cdot 10^3$ & 0.9 \\
		$2JAV \rightarrow 3JAV$ & 1.0 & $0.9 \cdot 10^3$ & 0.5\\
		\hline
	\end{tabularx}
\end{table*}
\end{singlespace}

The dynamics of hysteretic (JAV) and non-hysteretic (JV) entries and related energies are very different. In Fig.~\ref{fig:Fraun}(f) we present the magnetic field dependence of the energy dissipated in the junction over one tip oscillation period. For each  $H_{ext}$, this energy is calculated as $E_{dis} = \sum_{k=1}^{31} \int_0^{t_0}I_{nk}U_{k}dt$, where $t_0$ is the tip oscillation period, $I_{nk}$ and $U_k$ are the quasiparticle current and the voltage drop across $k$-th junction of the model. In the hysteretic regime, the dissipation is relatively high only during JAV entry events: $E_{dis} = 1.7$, $0.9$, $0.5$~aJ, for the transitions 0 $\longleftrightarrow 1$, $1 \longleftrightarrow 2$, and $2 \longleftrightarrow 3$, respectively. These events are visible as strong peaks on the red curve. JAV entry velocity is fast, $\sim 4.6\times10^3$~m/s, $\sim 1.73\times10^3$~m/s, and $\sim 0.89\times10^3$~m/s.  Though, these events occur only during the first oscillation period. Once inside the junction, the new $(n+1)$JAV configuration remains stable: The following tip oscillations and even slight variations of $H_{ext}$ cannot trigger JAV escape. This is clear from $E_{dis}(H_{ext})$ calculated for the following oscillation periods (black curve) where the peaks are absent. One can also see the peak amplitude  as the height of the potential barrier for JAV to overcome, $\sim$1~aJ~$\simeq 10^{5} k_BT$, making the thermally activated escape improbable.


In the non-hysteretic regime (realized for JVs at $H_{ext}<$~-50 Oe) the barrier is absent, and the JV entry/escape occurs repeatedly at each tip oscillation. When the tip approaches the junction, it pushes the JV out; when it retracts, it let them getting in. As a result, the curves $E_{dis}(H_{ext})$ for the first oscillation and the following ones overlap. 

We now focus on the region selected by yellow dashed square in Fig.~\ref{fig:Fraun}(a). There, the spatially inhomogeneous magnetic flux in the junction enables realizing very peculiar JV phase portraits. Indeed, by slightly varying $I$ and $H_{ext}$ it becomes possible to reach a Meissner state (0JV), 2JV state, and two independent 1JV states in-between. These 1JV states are realized due to the possibility for JVs to enter from either side of the JJ. By applying a negative current across the junction helps the JVs to enter from one side and prevents its entrance from the other side, and verse versa. Remarkably, these 4 phases share a common point. Starting from the Meissner phase and by crossing this point it becomes possible to reach 2JV phase directly. Topologically, this 0JV $\rightarrow$ 2JV transition corresponds to the simultaneous entry of two individual JVs from the opposite edges of the JJ. Fig.~\ref{fig:Fraun}(e) shows the calculated dependence of the derivative of the total flux in the JJ on the current and field for the same $(I,H_{ext})$ region. Definitely, our model nicely reproduces this unusual effect and confirms the existence of the crossing point.

\section{Discussion}

The discovery of a peculiar crossing point, the revealed striking differences in the behaviour of JV and JAV demonstrate the interest for realizing a spatially varying magnetic flux in long JJ devices. While in the present work it was achieved by superimposing the local field of the MFM tip with an external field of a coil, in advanced devices it could be done locally by putting small ferromagnetic elements\cite{Samokhvalov_2012} or trapped Abrikosov vortex\cite{Golod_2021} in desired locations near the junction. Moreover, instead of using an external coil, a supercurrent $I_{along}$ can be applied along  the superconducting electrodes  to produce an extra-flux inside the junction,  
and to drive it, for instance, into a fast and hysteretic JV entry/escape regime, as achieved in our experiment at -35 Oe $< H_{ext} <$ 0 Oe, Fig.~\ref{fig:Fraun}(a-c). In this regime, the JJ is a bi-stable system with an ability to have different numbers of JVs at the same external field, that is 0/1 at $H_{ext} = -32$~Oe, 1/2 at $H_{ext} = -17$~Oe, etc. This can be used in low-dissipative logic and/or memory elements with a high integration density which is a challenge \cite{Semenov_2019}. 

As a proof of principle, we present in Fig.~\ref{IPH-along} the experiment demonstrating $H_{ext} \leftrightarrow I_{along}$ interchangeability in the JJ control. Figs.~\ref{IPH-along}(a,b) are tip oscillation phase maps obtained at different currents $I_{along}$ and fields $H_{ext}$. They reveal the phase drops corresponding to JV transitions with a nearly linear $H_{ext}(I_{along})$ dependence $H_{ext}[$Oe$]\simeq-5~I_{along}[$mA$]$ and thus, the capacity to access, by varying $I_{along}$, all JV states even with no external magnetic field applied, Fig.~\ref{IPH-along}(c). The comparison between forward (a) and backward (b) swept currents demonstrate that the hysteresis effects observed in $H_{ext}$ sweeps, Fig.~\ref{fig:Fraun}(a-d)), are also reproduced in $I_{along}$ sweeps, Fig.~\ref{IPH-along}(a-c).

Fig.~\ref{IPH-along}d shows the variation of the magnetic flux inside the junction upon application of current pulses; this demonstrates the JJ use as a fast write/read memory device.
A short positive current pulse $I_{along}>$0 leads to the entry of one JV into contact (transition 0 $\rightarrow$ 1). A current pulse of the same polarity (applied at $t=$40 ns) does not change the JV state. A current pulse of the same amplitude but of the opposite sign (at $t=$60 ns) triggers the JV escape (1 $\rightarrow$ 0). The repetition of the negative pulse (at $t=$85 ns) does not modify the 0JV state, etc. It is important to note that in these simulations, the pulse amplitude was intentionally taken below the critical current value for the selected operation point. This enables one switching between JV states while keeping JJ in the zero-resistance regime. The switching speed is limited by the entry/escape time of JV. For the selected working point ($H_{ext}=-$33 Oe, $I_{along}$=0), the operation time is 0.43 ns, and the dissipated energy $E_{dis}=$ 1.8 aJ -- a low value for memory devices. The switching time and the dissipated energy for different JV states are summarized in the table I.

\section{Conclusion}
To summarize, in this work we have studied the Josephson vortex penetration and escape dynamics in long planar Nb-Cu-Nb proximity junctions. We demonstrated experimentally and confirmed theoretically that by generating a spatially inhomogeneous magnetic flux inside the junction, it becomes possible to create peculiar Josephson states and to switch between them while globally preserving the superconducting state of the junction. We suggest to employ these low-dissipative Josephson vortex states in all-supercurrent-controlled logic and memory devices with no need for an external field to be applied for flux manipulation. The devices are advantageously characterized by fast switching times, low dissipation and high integration densities.



\begin{acknowledgement}
Authors thank D. Baranov for the technical assistance and N. Bergeal for fruitful discussions.  MFM experiments were carried out with the support of the Russian Science Foundation (project No. 18-72-10118). Numerical modeling were carried out with the support of the Russian Science Foundation (project No. 20-69-47013). This work was performed using e-beam lithography of MIPT Shared Facilities Center,  with  financial  support  from  the  Ministry  of  Education  and  Science  of  the  Russian Federation. This work was partially supported by the Ministry of Science and Higher Education of the Russian Federation (No. FSMG-2021-0005)
\end{acknowledgement}

\textbf{Author contribution.}  V.S.S. suggested the idea of the experiment; V.S.S. conceived the project and supervised the experiments; V.S.S. provided the combined MFM/transport experiment with assistance from R.A.H., V.S.S. and A.G.Sh. realized e-beam lithography and thin film deposition; V.S.S., D.R., V.I.R, N.V.K., I.I.S, M.Yu.K. and A.V.A. provided the explanation of the observed phenomena; V.I.R, I.I.S, V.S.S. and D.R. constructed the model of the observed effects; V.I.R and I.I.S provided numerical simulations; D.R., V.S.S., V.I.R, I.I.S, and R.A.H. wrote the manuscript with the contributions from other authors.




\bibliography{Nano}

\providecommand{\latin}[1]{#1}
\makeatletter
\providecommand{\doi}
  {\begingroup\let\do\@makeother\dospecials
  \catcode`\{=1 \catcode`\}=2 \doi@aux}
\providecommand{\doi@aux}[1]{\endgroup\texttt{#1}}
\makeatother
\providecommand*\mcitethebibliography{\thebibliography}
\csname @ifundefined\endcsname{endmcitethebibliography}
  {\let\endmcitethebibliography\endthebibliography}{}
\begin{mcitethebibliography}{44}
\providecommand*\natexlab[1]{#1}
\providecommand*\mciteSetBstSublistMode[1]{}
\providecommand*\mciteSetBstMaxWidthForm[2]{}
\providecommand*\mciteBstWouldAddEndPuncttrue
  {\def\EndOfBibitem{\unskip.}}
\providecommand*\mciteBstWouldAddEndPunctfalse
  {\let\EndOfBibitem\relax}
\providecommand*\mciteSetBstMidEndSepPunct[3]{}
\providecommand*\mciteSetBstSublistLabelBeginEnd[3]{}
\providecommand*\EndOfBibitem{}
\mciteSetBstSublistMode{f}
\mciteSetBstMaxWidthForm{subitem}{(\alph{mcitesubitemcount})}
\mciteSetBstSublistLabelBeginEnd
  {\mcitemaxwidthsubitemform\space}
  {\relax}
  {\relax}

\bibitem[Andreev(1965)]{Andreev}
Andreev,~A. Thermal conductivity of the intermediate state of superconductors
  II. \emph{Sov. Phys. JETP} \textbf{1965}, \emph{20}, 1490\relax
\mciteBstWouldAddEndPuncttrue
\mciteSetBstMidEndSepPunct{\mcitedefaultmidpunct}
{\mcitedefaultendpunct}{\mcitedefaultseppunct}\relax
\EndOfBibitem
\bibitem[Abrikosov(1957)]{Abrikosov_1957}
Abrikosov,~A. The magnetic properties of superconducting alloys. \emph{Journal
  of Physics and Chemistry of Solids} \textbf{1957}, \emph{2}, 199--208\relax
\mciteBstWouldAddEndPuncttrue
\mciteSetBstMidEndSepPunct{\mcitedefaultmidpunct}
{\mcitedefaultendpunct}{\mcitedefaultseppunct}\relax
\EndOfBibitem
\bibitem[Stolyarov \latin{et~al.}(2018)Stolyarov, Cren, Brun, Golovchanskiy,
  Skryabina, Kasatonov, Khapaev, Kupriyanov, Golubov, and
  Roditchev]{Stolyarov_2018}
Stolyarov,~V.~S.; Cren,~T.; Brun,~C.; Golovchanskiy,~I.~A.; Skryabina,~O.~V.;
  Kasatonov,~D.~I.; Khapaev,~M.~M.; Kupriyanov,~M.~Y.; Golubov,~A.~A.;
  Roditchev,~D. Expansion of a superconducting vortex core into a diffusive
  metal. \emph{Nature communications} \textbf{2018}, \emph{9}, 1--8\relax
\mciteBstWouldAddEndPuncttrue
\mciteSetBstMidEndSepPunct{\mcitedefaultmidpunct}
{\mcitedefaultendpunct}{\mcitedefaultseppunct}\relax
\EndOfBibitem
\bibitem[Josephson(1962)]{Josephson_1962}
Josephson,~B.~D. Possible new effects in superconductive tunnelling.
  \emph{Physics letters} \textbf{1962}, \emph{1}, 251--253\relax
\mciteBstWouldAddEndPuncttrue
\mciteSetBstMidEndSepPunct{\mcitedefaultmidpunct}
{\mcitedefaultendpunct}{\mcitedefaultseppunct}\relax
\EndOfBibitem
\bibitem[Anderson and Rowell(1963)Anderson, and Rowell]{Anderson_1963}
Anderson,~P.~W.; Rowell,~J.~M. Probable observation of the Josephson
  superconducting tunneling effect. \emph{Physical Review Letters}
  \textbf{1963}, \emph{10}, 230\relax
\mciteBstWouldAddEndPuncttrue
\mciteSetBstMidEndSepPunct{\mcitedefaultmidpunct}
{\mcitedefaultendpunct}{\mcitedefaultseppunct}\relax
\EndOfBibitem
\bibitem[Rowell(1963)]{Rowell_1963}
Rowell,~J. Magnetic field dependence of the Josephson tunnel current.
  \emph{Physical Review Letters} \textbf{1963}, \emph{11}, 200\relax
\mciteBstWouldAddEndPuncttrue
\mciteSetBstMidEndSepPunct{\mcitedefaultmidpunct}
{\mcitedefaultendpunct}{\mcitedefaultseppunct}\relax
\EndOfBibitem
\bibitem[Josephson(1965)]{Josephson_1965}
Josephson,~B.~D. Supercurrents through barriers. \emph{Advances in Physics}
  \textbf{1965}, \emph{14}, 419--451\relax
\mciteBstWouldAddEndPuncttrue
\mciteSetBstMidEndSepPunct{\mcitedefaultmidpunct}
{\mcitedefaultendpunct}{\mcitedefaultseppunct}\relax
\EndOfBibitem
\bibitem[Wallraff \latin{et~al.}(2003)Wallraff, Lukashenko, Lisenfeld, Kemp,
  Fistul, Koval, and Ustinov]{Wallraff_2003}
Wallraff,~A.; Lukashenko,~A.; Lisenfeld,~J.; Kemp,~A.; Fistul,~M.; Koval,~Y.;
  Ustinov,~A. Quantum dynamics of a single vortex. \emph{Nature} \textbf{2003},
  \emph{425}, 155--158\relax
\mciteBstWouldAddEndPuncttrue
\mciteSetBstMidEndSepPunct{\mcitedefaultmidpunct}
{\mcitedefaultendpunct}{\mcitedefaultseppunct}\relax
\EndOfBibitem
\bibitem[Berdiyorov \latin{et~al.}(2018)Berdiyorov, Milo{\v{s}}evi{\'c},
  Kusmartsev, Peeters, and Savel’ev]{Berdiyorov_2018}
Berdiyorov,~G.; Milo{\v{s}}evi{\'c},~M.; Kusmartsev,~F.; Peeters,~F.;
  Savel’ev,~S. Josephson vortex loops in nanostructured Josephson junctions.
  \emph{Scientific reports} \textbf{2018}, \emph{8}, 1--17\relax
\mciteBstWouldAddEndPuncttrue
\mciteSetBstMidEndSepPunct{\mcitedefaultmidpunct}
{\mcitedefaultendpunct}{\mcitedefaultseppunct}\relax
\EndOfBibitem
\bibitem[Cuevas and Bergeret(2007)Cuevas, and Bergeret]{Cuevas_2007}
Cuevas,~J.; Bergeret,~F. Magnetic interference patterns and vortices in
  diffusive SNS junctions. \emph{Physical review letters} \textbf{2007},
  \emph{99}, 217002\relax
\mciteBstWouldAddEndPuncttrue
\mciteSetBstMidEndSepPunct{\mcitedefaultmidpunct}
{\mcitedefaultendpunct}{\mcitedefaultseppunct}\relax
\EndOfBibitem
\bibitem[Roditchev \latin{et~al.}(2015)Roditchev, Brun, Serrier-Garcia, Cuevas,
  Bessa, Milo{\v{s}}evi{\'c}, Debontridder, Stolyarov, and
  Cren]{Roditchev_2015}
Roditchev,~D.; Brun,~C.; Serrier-Garcia,~L.; Cuevas,~J.~C.; Bessa,~V. H.~L.;
  Milo{\v{s}}evi{\'c},~M.~V.; Debontridder,~F.; Stolyarov,~V.; Cren,~T. Direct
  observation of Josephson vortex cores. \emph{Nature Physics} \textbf{2015},
  \emph{11}, 332\relax
\mciteBstWouldAddEndPuncttrue
\mciteSetBstMidEndSepPunct{\mcitedefaultmidpunct}
{\mcitedefaultendpunct}{\mcitedefaultseppunct}\relax
\EndOfBibitem
\bibitem[Dremov \latin{et~al.}(2019)Dremov, Grebenchuk, Shishkin, Baranov,
  Hovhannisyan, Skryabina, Golovchanskiy, Chichkov, Brun, Cren, \latin{et~al.}
  others]{Dremov_2019}
Dremov,~V.~V.; Grebenchuk,~S.~Y.; Shishkin,~A.~G.; Baranov,~D.~S.;
  Hovhannisyan,~R.~A.; Skryabina,~O.~V.; Golovchanskiy,~I.~A.; Chichkov,~V.~I.;
  Brun,~C.; Cren,~T., \latin{et~al.}  Local Josephson vortex generation and
  manipulation with a Magnetic Force Microscope. \emph{Nature communications}
  \textbf{2019}, \emph{10}, 1--9\relax
\mciteBstWouldAddEndPuncttrue
\mciteSetBstMidEndSepPunct{\mcitedefaultmidpunct}
{\mcitedefaultendpunct}{\mcitedefaultseppunct}\relax
\EndOfBibitem
\bibitem[Grebenchuk \latin{et~al.}(2020)Grebenchuk, Hovhannisyan, Dremov,
  Shishkin, Chichkov, Golubov, Roditchev, Krasnov, and
  Stolyarov]{Grbenchuk_2020}
Grebenchuk,~S.~Y.; Hovhannisyan,~R.~A.; Dremov,~V.~V.; Shishkin,~A.~G.;
  Chichkov,~V.~I.; Golubov,~A.~A.; Roditchev,~D.; Krasnov,~V.~M.;
  Stolyarov,~V.~S. Observation of interacting Josephson vortex chains by
  magnetic force microscopy. \emph{Phys. Rev. Research} \textbf{2020},
  \emph{2}, 023105\relax
\mciteBstWouldAddEndPuncttrue
\mciteSetBstMidEndSepPunct{\mcitedefaultmidpunct}
{\mcitedefaultendpunct}{\mcitedefaultseppunct}\relax
\EndOfBibitem
\bibitem[Hovhannisyan \latin{et~al.}(2021)Hovhannisyan, Grebenchuk, Baranov,
  Roditchev, and Stolyarov]{Hovhannisyan_2021}
Hovhannisyan,~R.~A.; Grebenchuk,~S.~Y.; Baranov,~D.~S.; Roditchev,~D.;
  Stolyarov,~V.~S. Lateral Josephson Junctions as Sensors for Magnetic
  Microscopy at Nanoscale. \emph{The Journal of Physical Chemistry Letters}
  \textbf{2021}, \emph{12}, 12196--12201\relax
\mciteBstWouldAddEndPuncttrue
\mciteSetBstMidEndSepPunct{\mcitedefaultmidpunct}
{\mcitedefaultendpunct}{\mcitedefaultseppunct}\relax
\EndOfBibitem
\bibitem[Krasnov(2020)]{Krasnov_2020}
Krasnov,~V.~M. Josephson junctions in a local inhomogeneous magnetic field.
  \emph{Physical Review B} \textbf{2020}, \emph{101}, 144507\relax
\mciteBstWouldAddEndPuncttrue
\mciteSetBstMidEndSepPunct{\mcitedefaultmidpunct}
{\mcitedefaultendpunct}{\mcitedefaultseppunct}\relax
\EndOfBibitem
\bibitem[Malomed and Ustinov(2004)Malomed, and Ustinov]{Malomed_2004}
Malomed,~B.~A.; Ustinov,~A.~V. Creation of classical and quantum fluxons by a
  current dipole in a long Josephson junction. \emph{Physical Review B}
  \textbf{2004}, \emph{69}, 064502\relax
\mciteBstWouldAddEndPuncttrue
\mciteSetBstMidEndSepPunct{\mcitedefaultmidpunct}
{\mcitedefaultendpunct}{\mcitedefaultseppunct}\relax
\EndOfBibitem
\bibitem[Gulevich and Kusmartsev(2006)Gulevich, and Kusmartsev]{Gulevich_2006}
Gulevich,~D.~R.; Kusmartsev,~F. Flux Cloning in Josephson Transmission Lines.
  \emph{Physical review letters} \textbf{2006}, \emph{97}, 017004\relax
\mciteBstWouldAddEndPuncttrue
\mciteSetBstMidEndSepPunct{\mcitedefaultmidpunct}
{\mcitedefaultendpunct}{\mcitedefaultseppunct}\relax
\EndOfBibitem
\bibitem[Gulevich \latin{et~al.}(2017)Gulevich, Koshelets, and
  Kusmartsev]{Gulevich_2017}
Gulevich,~D.~R.; Koshelets,~V.~P.; Kusmartsev,~F.~V. Josephson flux-flow
  oscillator: The microscopic tunneling approach. \emph{Physical Review B}
  \textbf{2017}, \emph{96}, 024515\relax
\mciteBstWouldAddEndPuncttrue
\mciteSetBstMidEndSepPunct{\mcitedefaultmidpunct}
{\mcitedefaultendpunct}{\mcitedefaultseppunct}\relax
\EndOfBibitem
\bibitem[Knufinke \latin{et~al.}(2012)Knufinke, K.~Ilin, D.~Koelle, and
  Goldobin]{Knufinke_2012}
Knufinke,~M.; K.~Ilin,~M.~S.; D.~Koelle,~R.~K.; Goldobin,~E. Deterministic
  Josephson vortex ratchet with a load. \emph{Physical Review E} \textbf{2012},
  \emph{85}, 011122\relax
\mciteBstWouldAddEndPuncttrue
\mciteSetBstMidEndSepPunct{\mcitedefaultmidpunct}
{\mcitedefaultendpunct}{\mcitedefaultseppunct}\relax
\EndOfBibitem
\bibitem[Orlando \latin{et~al.}(1999)Orlando, Mooij, Tian, van~der Wal,
  Levitov, Lloyd, and Mazo]{FlQ1}
Orlando,~T.~P.; Mooij,~J.~E.; Tian,~L.; van~der Wal,~C.~H.; Levitov,~L.~S.;
  Lloyd,~S.; Mazo,~J.~J. Superconducting persistent-current qubit. \emph{Phys.
  Rev. B} \textbf{1999}, \emph{60}, 15398\relax
\mciteBstWouldAddEndPuncttrue
\mciteSetBstMidEndSepPunct{\mcitedefaultmidpunct}
{\mcitedefaultendpunct}{\mcitedefaultseppunct}\relax
\EndOfBibitem
\bibitem[Friedman \latin{et~al.}(2000)Friedman, Patel, Chen, Tolpygo, and
  Lukens]{FlQ2}
Friedman,~J.~R.; Patel,~V.; Chen,~W.; Tolpygo,~S.~K.; Lukens,~J.~E. Quantum
  superposition of distinct macroscopic states. \emph{Nature} \textbf{2000},
  \emph{406}, 43--46\relax
\mciteBstWouldAddEndPuncttrue
\mciteSetBstMidEndSepPunct{\mcitedefaultmidpunct}
{\mcitedefaultendpunct}{\mcitedefaultseppunct}\relax
\EndOfBibitem
\bibitem[Kemp \latin{et~al.}(2002)Kemp, Wallraff, and Ustinov]{JVQ1}
Kemp,~A.; Wallraff,~A.; Ustinov,~A.~V. Josephson Vortex Qubit: Design,
  Preparation and Read-Out. \emph{Phys. Status Solidi B} \textbf{2002},
  \emph{233}, 472\relax
\mciteBstWouldAddEndPuncttrue
\mciteSetBstMidEndSepPunct{\mcitedefaultmidpunct}
{\mcitedefaultendpunct}{\mcitedefaultseppunct}\relax
\EndOfBibitem
\bibitem[Clarke(2003)]{JVQ2}
Clarke,~J. Vortices and hearts. \emph{Nature} \textbf{2003}, \emph{425},
  133\relax
\mciteBstWouldAddEndPuncttrue
\mciteSetBstMidEndSepPunct{\mcitedefaultmidpunct}
{\mcitedefaultendpunct}{\mcitedefaultseppunct}\relax
\EndOfBibitem
\bibitem[McDermott \latin{et~al.}(2018)McDermott, Vavilov, Plourde, Wilhelm,
  Liebermann, Mukhanov, and Ohki]{QuProcMukh}
McDermott,~R.; Vavilov,~M.; Plourde,~B. L.~T.; Wilhelm,~F.~K.;
  Liebermann,~P.~J.; Mukhanov,~O.~A.; Ohki,~T.~A. Quantum–classical interface
  based on single flux quantum digital logic. \emph{Quantum Sci. Technol.}
  \textbf{2018}, \emph{3}, 024004\relax
\mciteBstWouldAddEndPuncttrue
\mciteSetBstMidEndSepPunct{\mcitedefaultmidpunct}
{\mcitedefaultendpunct}{\mcitedefaultseppunct}\relax
\EndOfBibitem
\bibitem[Soloviev \latin{et~al.}(2015)Soloviev, Klenov, Pankratov, Revin,
  Il’ichev, and Kuzmin]{SolScat}
Soloviev,~I.~I.; Klenov,~N.~V.; Pankratov,~A.~L.; Revin,~L.~S.; Il’ichev,~E.;
  Kuzmin,~L.~S. Soliton scattering as a measurement tool for weak signals.
  \emph{Phys. Rev. B} \textbf{2015}, \emph{92}, 014516\relax
\mciteBstWouldAddEndPuncttrue
\mciteSetBstMidEndSepPunct{\mcitedefaultmidpunct}
{\mcitedefaultendpunct}{\mcitedefaultseppunct}\relax
\EndOfBibitem
\bibitem[Soloviev \latin{et~al.}(2014)Soloviev, Klenov, Bakurskiy, Pankratov,
  and Kuzmin]{BalRoc}
Soloviev,~I.~I.; Klenov,~N.~V.; Bakurskiy,~S.~V.; Pankratov,~A.~L.;
  Kuzmin,~L.~S. Symmetrical Josephson vortex interferometer as an advanced
  ballistic single-shot detector. \emph{Appl. Phys. Lett.} \textbf{2014},
  \emph{105}, 202602\relax
\mciteBstWouldAddEndPuncttrue
\mciteSetBstMidEndSepPunct{\mcitedefaultmidpunct}
{\mcitedefaultendpunct}{\mcitedefaultseppunct}\relax
\EndOfBibitem
\bibitem[Ishida \latin{et~al.}(2021)Ishida, Byun, Nagaoka, Fukumitsu, Tanaka,
  Kawakami, Tanimoto, Ono, Kim, and Inoue]{SCNN1}
Ishida,~K.; Byun,~I.; Nagaoka,~I.; Fukumitsu,~K.; Tanaka,~M.; Kawakami,~S.;
  Tanimoto,~T.; Ono,~T.; Kim,~J.; Inoue,~K. Superconductor Computing for Neural
  Networks. \emph{IEEE Micro} \textbf{2021}, \emph{41}, 19--26\relax
\mciteBstWouldAddEndPuncttrue
\mciteSetBstMidEndSepPunct{\mcitedefaultmidpunct}
{\mcitedefaultendpunct}{\mcitedefaultseppunct}\relax
\EndOfBibitem
\bibitem[Cai \latin{et~al.}(2019)Cai, Ren, Chen, Liu, Ding, Qian, Han, Luo,
  Yoshikawa, and Wang]{SCNN2}
Cai,~R.; Ren,~A.; Chen,~O.; Liu,~N.; Ding,~C.; Qian,~X.; Han,~J.; Luo,~W.;
  Yoshikawa,~N.; Wang,~Y. A Stochastic-Computing Based Deep Learning Framework
  Using Adiabatic Quantum-Flux-Parametron Superconducting Technology.
  Proceedings of the 46th International Symposium on Computer Architecture. New
  York, NY, USA, 2019; p 567–578\relax
\mciteBstWouldAddEndPuncttrue
\mciteSetBstMidEndSepPunct{\mcitedefaultmidpunct}
{\mcitedefaultendpunct}{\mcitedefaultseppunct}\relax
\EndOfBibitem
\bibitem[Soloviev \latin{et~al.}(2018)Soloviev, Schegolev, Klenov, Bakurskiy,
  Kupriyanov, Tereshonok, Shadrin, Stolyarov, and Golubov]{SCNN3}
Soloviev,~I.~I.; Schegolev,~A.~E.; Klenov,~N.~V.; Bakurskiy,~S.~V.;
  Kupriyanov,~M.~Y.; Tereshonok,~M.~V.; Shadrin,~A.~V.; Stolyarov,~V.~S.;
  Golubov,~A.~A. Adiabatic superconducting artificial neural network: Basic
  cells. \emph{J. Appl. Phys.} \textbf{2018}, \emph{124}, 152113\relax
\mciteBstWouldAddEndPuncttrue
\mciteSetBstMidEndSepPunct{\mcitedefaultmidpunct}
{\mcitedefaultendpunct}{\mcitedefaultseppunct}\relax
\EndOfBibitem
\bibitem[Rowlands \latin{et~al.}(2021)Rowlands, Nguyen, Ribeill, Wagner, Govia,
  Barbosa, Gauthier, and Ohki]{ResC}
Rowlands,~G.~E.; Nguyen,~M.-H.; Ribeill,~G.~J.; Wagner,~A.~P.; Govia,~L. C.~G.;
  Barbosa,~W. A.~S.; Gauthier,~D.~J.; Ohki,~T.~A. Reservoir Computing with
  Superconducting Electronics. \emph{arXiv:2103.02522} \textbf{2021}, \relax
\mciteBstWouldAddEndPunctfalse
\mciteSetBstMidEndSepPunct{\mcitedefaultmidpunct}
{}{\mcitedefaultseppunct}\relax
\EndOfBibitem
\bibitem[Holmes \latin{et~al.}(2013)Holmes, Ripple, and Manheimer]{Holmes}
Holmes,~D.~S.; Ripple,~A.~L.; Manheimer,~M.~A. Energy-Efficient Superconducting
  Computing-Power Budgets and Requirements. \emph{IEEE Trans. Appl. Supercond.}
  \textbf{2013}, \emph{23}, 1701610\relax
\mciteBstWouldAddEndPuncttrue
\mciteSetBstMidEndSepPunct{\mcitedefaultmidpunct}
{\mcitedefaultendpunct}{\mcitedefaultseppunct}\relax
\EndOfBibitem
\bibitem[Mukhanov(2015)]{Mukh2}
Mukhanov,~O.~A. In \emph{Applied Superconductivity: Handbook on Devices and
  Applications}; Seidel,~P., Ed.; Wiley-VCH Verlag GmbH \& Co. KGaA: Weinheim,
  Germany, 2015; Chapter Superconductor Digital Electronics, pp 1--28\relax
\mciteBstWouldAddEndPuncttrue
\mciteSetBstMidEndSepPunct{\mcitedefaultmidpunct}
{\mcitedefaultendpunct}{\mcitedefaultseppunct}\relax
\EndOfBibitem
\bibitem[Tolpygo(2016)]{Tolp}
Tolpygo,~S.~K. Superconductor Digital Electronics: Scalability and Energy
  Efficiency Issues. \emph{Low Temp. Phys.} \textbf{2016}, \emph{42}, 361\relax
\mciteBstWouldAddEndPuncttrue
\mciteSetBstMidEndSepPunct{\mcitedefaultmidpunct}
{\mcitedefaultendpunct}{\mcitedefaultseppunct}\relax
\EndOfBibitem
\bibitem[Soloviev \latin{et~al.}(2017)Soloviev, Klenov, Bakurskiy, Kupriyanov,
  Gudkov, and Sidorenko]{Beil}
Soloviev,~I.~I.; Klenov,~N.~V.; Bakurskiy,~S.~V.; Kupriyanov,~M.~Y.;
  Gudkov,~A.~L.; Sidorenko,~A.~S. Beyond {M}oore's technologies: operation
  principles of a superconductor alternative. \emph{Beilstein J. Nanotechnol}
  \textbf{2017}, \emph{8}, 2689--2710\relax
\mciteBstWouldAddEndPuncttrue
\mciteSetBstMidEndSepPunct{\mcitedefaultmidpunct}
{\mcitedefaultendpunct}{\mcitedefaultseppunct}\relax
\EndOfBibitem
\bibitem[Bhushan \latin{et~al.}(2020)Bhushan, Bunyk, Cuthbert, DeBenedictis,
  Fagaly, Febvre, Fourie, Frank, Gupta, Herr, Holmes, Humble, de~Escobar,
  McGeoch, Missert, Mueller, Mukhanov, Nemoto, Rao, Patra, Plourde, Pugach,
  Tyrrell, Vogelsang, Wilhelm-Mauch, and Yoshikawa]{IRDS2020}
Bhushan,~M. \latin{et~al.}  {IRDS} 2020: {C}ryogenic {E}lectronics and
  {Q}uantum {I}nformation {P}rocessing. https://irds.ieee.org/editions/2020,
  2020; Part of {IEEE} {I}nternational {R}oadmap for {D}evices and
  {S}ystems\relax
\mciteBstWouldAddEndPuncttrue
\mciteSetBstMidEndSepPunct{\mcitedefaultmidpunct}
{\mcitedefaultendpunct}{\mcitedefaultseppunct}\relax
\EndOfBibitem
\bibitem[Semenov \latin{et~al.}(2019)Semenov, Polyakov, and
  Tolpygo]{Semenov_2019}
Semenov,~V.~K.; Polyakov,~Y.~A.; Tolpygo,~S.~K. Very large scale integration of
  Josephson-junction-based superconductor random access memories. \emph{IEEE
  Transactions on Applied Superconductivity} \textbf{2019}, \emph{29},
  1--9\relax
\mciteBstWouldAddEndPuncttrue
\mciteSetBstMidEndSepPunct{\mcitedefaultmidpunct}
{\mcitedefaultendpunct}{\mcitedefaultseppunct}\relax
\EndOfBibitem
\bibitem[Ilin \latin{et~al.}(2021)Ilin, Song, Burkova, Silge, Z.~Guo, and
  Bezryadina]{Ilin_2021}
Ilin,~E.; Song,~X.; Burkova,~I.; Silge,~A.; Z.~Guo,~K.~I.; Bezryadina,~A.
  Supercurrent-controlled kinetic inductance superconducting memory element.
  \emph{Applied Physics Letters} \textbf{2021}, \emph{118}, 112603\relax
\mciteBstWouldAddEndPuncttrue
\mciteSetBstMidEndSepPunct{\mcitedefaultmidpunct}
{\mcitedefaultendpunct}{\mcitedefaultseppunct}\relax
\EndOfBibitem
\bibitem[Miloshevsky \latin{et~al.}(2022)Miloshevsky, Nair, Imam, and
  Braiman]{Miloshevsky_2022}
Miloshevsky,~A.; Nair,~N.; Imam,~N.; Braiman,~Y. High‑Tc Superconducting
  Memory Cell. \emph{Journal of Superconductivity and Novel Magnetism}
  \textbf{2022}, \emph{35}, 373–--382\relax
\mciteBstWouldAddEndPuncttrue
\mciteSetBstMidEndSepPunct{\mcitedefaultmidpunct}
{\mcitedefaultendpunct}{\mcitedefaultseppunct}\relax
\EndOfBibitem
\bibitem[Ligato \latin{et~al.}(2021)Ligato, Strambini, Paolucci, and
  Giazotto]{Ligato_2021}
Ligato,~N.; Strambini,~E.; Paolucci,~F.; Giazotto,~F. Preliminary demonstration
  of a persistent Josephson phase-slip memory cell with topological protection.
  \emph{Nature communications} \textbf{2021}, \emph{12}\relax
\mciteBstWouldAddEndPuncttrue
\mciteSetBstMidEndSepPunct{\mcitedefaultmidpunct}
{\mcitedefaultendpunct}{\mcitedefaultseppunct}\relax
\EndOfBibitem
\bibitem[Ruzhitskiy \latin{et~al.}(2021)Ruzhitskiy, Soloviev, Bakurskiy, and
  et~al.]{Ruzhitskiy_2021}
Ruzhitskiy,~V.; Soloviev,~I.; Bakurskiy,~S.; et~al., Modeling of the vortex
  dynamics in long josephson junction. \emph{2021 IEEE 14th Workshop on Low
  Temperature Electronics (WOLTE)} \textbf{2021}, \relax
\mciteBstWouldAddEndPunctfalse
\mciteSetBstMidEndSepPunct{\mcitedefaultmidpunct}
{}{\mcitedefaultseppunct}\relax
\EndOfBibitem
\bibitem[Golovchanskiy \latin{et~al.}(2017)Golovchanskiy, Abramov, Stolyarov,
  Emelyanova, Golubov, Ustinov, and Ryazanov]{Golovchanskiy_SUST_30_054005}
Golovchanskiy,~I.~A.; Abramov,~N.~N.; Stolyarov,~V.~S.; Emelyanova,~O.~V.;
  Golubov,~A.~A.; Ustinov,~A.~V.; Ryazanov,~V.~V. Ferromagnetic resonance with
  long Josephson junction. \emph{Supercond. Sci. Technol.} \textbf{2017},
  \emph{30}, 054005\relax
\mciteBstWouldAddEndPuncttrue
\mciteSetBstMidEndSepPunct{\mcitedefaultmidpunct}
{\mcitedefaultendpunct}{\mcitedefaultseppunct}\relax
\EndOfBibitem
\bibitem[Samokhvalov \latin{et~al.}(2012)Samokhvalov, Vdovichev, Gribkov,
  Gusev, Klimov, Nozdrin, Rogov, Fraerman, Egorov, Bol'ginov, \latin{et~al.}
  others]{Samokhvalov_2012}
Samokhvalov,~A.; Vdovichev,~S.; Gribkov,~B.; Gusev,~S.; Klimov,~A.;
  Nozdrin,~Y.; Rogov,~V.; Fraerman,~A.; Egorov,~S.; Bol'ginov,~V.,
  \latin{et~al.}  Properties of Josephson junctions in the nonuniform field of
  ferromagnetic particles. \emph{JETP letters} \textbf{2012}, \emph{95}\relax
\mciteBstWouldAddEndPuncttrue
\mciteSetBstMidEndSepPunct{\mcitedefaultmidpunct}
{\mcitedefaultendpunct}{\mcitedefaultseppunct}\relax
\EndOfBibitem
\bibitem[Golod \latin{et~al.}(2021)Golod, Hovhannisyan, Kapran, Dremov,
  Stolyarov, and Krasnov]{Golod_2021}
Golod,~T.; Hovhannisyan,~R.~A.; Kapran,~O.~M.; Dremov,~V.~V.; Stolyarov,~V.~S.;
  Krasnov,~V.~M. Reconfigurable Josephson Phase Shifter. \emph{Nano Letters}
  \textbf{2021}, \emph{21}, 5240--5246\relax
\mciteBstWouldAddEndPuncttrue
\mciteSetBstMidEndSepPunct{\mcitedefaultmidpunct}
{\mcitedefaultendpunct}{\mcitedefaultseppunct}\relax
\EndOfBibitem
\end{mcitethebibliography}

\end{document}